\documentclass[superscriptaddress,twocolumn,showpacs,
amssymb,amsmath,nobibnotes,aps,prd,%
nofootinbib]{revtex4-1}
\pdfoutput=1
\usepackage{graphicx,bm,color,psfrag,hyperref,esvect}
\usepackage{cleveref}
\usepackage{amsfonts}
\usepackage{lipsum}
\usepackage{mathtools}
\usepackage{subcaption}
\usepackage{verbatim}
\usepackage[normalem]{ulem}
\usepackage[dvipsnames]{xcolor}
\usepackage{tabularx, comment, booktabs}

\hypersetup{colorlinks,linkcolor={blue},citecolor={red},urlcolor={violet}}

\newcommand{\C}{\textcolor{blue}}

\begin{document}

\preprint{APS/123-QED}

\title{Microlensing Black Hole Shadows-II: Constraining Primordial Black Hole Dark Matter using the photon rings of M87 and Sgr A*}

\author{Himanshu Verma}
\email{hverma@lsu.edu}
\affiliation{Department of Physics and Astronomy, Louisiana State University, Baton Rouge, LA 70803, USA}

\author{Priyanka Sarmah}
\email{sarmahpriyanka07@gmail.com}
\affiliation{Department of Physics, National Tsing Hua University, Hsinchu 30013, Taiwan}
\affiliation{Center for Theory and Computation, National Tsing Hua University, Hsinchu 30013, Taiwan}

\author{Joseph Silk}
\email{silk@iap.fr}
\affiliation{Institut d'Astrophysique de Paris (UMR7095: CNRS \& UPMC- Sorbonne Universities), F-75014, Paris, France}
\affiliation{William H. Miller III Department of Physics and Astronomy, The Johns Hopkins University, Baltimore, MD 21218, USA}
\affiliation{BIPAC, Department of Physics, University of Oxford, Keble Road, Oxford OX1 3RH, UK}

\author{Kingman Cheung}
\email{cheung@phys.nthu.edu.tw}
\affiliation{Department of Physics, National Tsing Hua University, Hsinchu 30013, Taiwan}
\affiliation{Center for Theory and Computation, National Tsing Hua University, Hsinchu 30013, Taiwan}
\affiliation{Division of Quantum Phases and Devices, School of Physics, Konkuk University, Seoul 143-701, Republic of Korea}

\date{\today}

\begin{abstract}
The resolution of photon rings of Sgr~A$^*$ and M87 is the next milestone of upcoming EHT-like interferometries. We extend the formalism developed in our previous work~\cite{Verma:2023hes} to constrain primordial black hole (PBH) dark matter using microlensing-induced distortions of black hole shadows. Building upon the theoretical framework for microlensing of photon rings, we apply this methodology to both Sgr A* and M87, considering multiple PBH populations: (i) PBH dark matter spikes around central supermassive black holes, (ii) NFW halo contributions in the Milky Way and M87 galaxies, and (iii) foreground Milky Way PBH dark matter affecting M87* observations. The microlensing signal manifests as a time-dependent asymmetry and deformation of the photon ring, providing the most sensitive observable for lensing effects. We assess the detectability of these signatures with future EHT-like interferometers. Our analysis reveals that M87* provides the strongest constraints on PBH dark matter. We show that the absence of photon-ring asymmetries in observations with angular resolution of order $0.1\,\mu{\rm as}$ can constrain PBHs in the mass range $10^{-5}\,M_\odot \lesssim M_{\rm PBH} \lesssim 10^{6}\,M_\odot$, with maximal sensitivity near $M_{\rm PBH}\sim10^{3}\,M_\odot$, for PBH dark matter fractions as small as $f_{\rm PBH}\sim10^{-2}$. 

\end{abstract}

\maketitle

\section{Introduction}
The nature of Dark matter (DM) remains elusive, persisting nearly a century after its first observational evidence in a galaxy cluster by Zwicky in 1933~\cite{Zwicky:1933gu, Zwicky:1937zza} and later in galactic rotation curves by Vera Rubin in the 1960s~\cite{vcrubin}. Despite comprising $\sim85\%$ of the matter in the universe, DM has revealed itself only through gravitational interactions. Among the diverse landscape of DM candidates, primordial black holes (PBHs) stand out as one of the leading ones. Unlike other candidates that require beyond standard model physics, PBHs can form naturally from the gravitational collapse of primordial density fluctuations in the early universe and could span a wide range from the Planck mass ($10^{-19}~M_\odot$) to stellar scales ($10^{4}~M_\odot$)~\cite{Hawking:1971ei, Clesse:2015wea, Garcia-Bellido:2017fdg, Carr:2020xqk, Choi:2022btl}.

Recent breakthroughs in very long baseline interferometry (VLBI) have revolutionized our ability to directly image the immediate vicinity of supermassive black holes (SMBHs). The Event Horizon Telescope (EHT) has successfully resolved the shadow of M87$^*$ and Sgr~A$^*$, providing unprecedented views of strong-field gravity regimes~\cite{EventHorizonTelescope:2019dse, EventHorizonTelescope:2019uob, EventHorizonTelescope:2019jan, EventHorizonTelescope:2019ths, EventHorizonTelescope:2019pgp, EventHorizonTelescope:2019ggy, EHT_sgra1, EHT_sgra2, EHT_sgra3}. These observations open new avenues for testing the PBH hypothesis through gravitational microlensing effects in these black hole images.

In the previous work~\cite{Verma:2023hes}, we developed a comprehensive formalism for microlensing signatures in black hole shadow observations. We demonstrated that stellar lenses in the Milky Way produce event rates $<10^{-3}$~yr$^{-1}$ for Sgr~A$^*$, making such effects challenging to observe with current or even futuristic baselines in space. However, the situation may be dramatically different for PBH dark matter, which can provide significantly larger optical depths due to higher number densities and extended spatial distributions.

The key insight is that microlensing affects not only the overall shadow morphology but also introduces characteristic time-dependent distortions in the photon ring structure. Unlike the Bardeen shadow (the theoretical inner boundary of trapped photon orbits), the observationally relevant shadow corresponds to the central brightened depression in EHT images. Microlensing signature will be a temporal variation in size, shape, and location of this brightness depression, with the photon ring serving as the most sensitive probe of such effects.

In this paper, we propose the use of photon ring movies of Sgr~A$^*$ and M87 to detect the existence of PBHs in the universe. These photon rings are expected to be captured with upcoming/proposed EHT-like interferometric facilities with baselines such as earth size: Next generation EHT (ngEHT)~\cite{2022Galax..10..111T, 2024ApJ...970L..24S}, earth-space: Black Hole Explorer (BHEX)~\cite{2024SPIE13092E..6QL, 2024SPIE13092E..2EA, 2024SPIE13092E..2DJ}, and space-space: Event Horizon Imager (EHI)~\cite{2019A&A...625A.124R, 2021ChJSS..41..211K, 2024A&A...686A.154S} (for a detailed review see~\cite{2024arXiv241201904B}). 

We consider the following scenarios that arise when targeting the two EHT sources:
\begin{itemize}
    \item \textbf{Sgr~A$^*$ scenarios:} (1) PBH spike microlensing of the Sgr~A$^*$ photon ring, and (2) Milky Way NFW population microlensing of Sgr~A$^*$.
    \item \textbf{M87$^*$ scenarios:} (3) PBH spike microlensing of the M87$^*$ photon ring, (4) M87 NFW PBH microlensing of M87$^*$, and (5) Milky Way NFW population mocrolensing of M87$^*$.
\end{itemize}


Although to the date of writing this manuscript, conclusive discoveries of PBHs are yet to be made, there have been various attempts to discover/constrain the average mass of PBHs and a fraction of dark matter in PBHs \cite{Carr:2020gox}. Microlensing of stars by PBHs places one of the most robust constraints. Photometric microlensing has already ruled out PBHs in the mass range $10^{-12}~M_\odot$ to $100~M_\odot$ and up to 1\% of DM by various microlensing surveys such as OGLE \cite{OGLE2019}, EROS \cite{EROS-2:2006ryy}, MACHO\cite{MACHO2000}, and HSC-Subaru\cite{HSC2019}. It has been speculated that the gravitational wave observation by the merger of the black holes in the mass range (5-100 $M_\odot$) may contain some of the events from PBHs\cite{gwbounds_Raidal:2017mfl}.

Observation of the photon ring, which also forms an outer envelope of Bardeen's shadow or the critical curves, remains an active direction of efforts by the EHT collaboration. Such observation is essential to fully set precise constraints on the spin and the mass of the SMBH at the center, which maximally affect the shape and the size of the photon ring. It is expected that with the ngEHT, the photon ring could be observable. In this paper, we study how the observation of a photon ring will open a new platform to search for a rather speculative population of PBHs as a candidate for dark matter. In ref.~\cite{Verma:2023hes}, it has been discussed that microlensing produces a time-varying modulation in the shape of the photon ring. Thus, in this paper, we calculate the event rate due to PBHs in EHT-like interferometry.

Previously, Ref.~\cite{Verma:2023hes} presents the formalism for the microlensing of the directly observable shadow. They apply the framework of microlensing to each point source of the boundary of the shadow due to a point lens in the foreground. Varying angular separations of the boundary sources from the lens lead to the apparent brightening and angular shift of each source on the boundary. This will be manifest as an overall distortion in the shadow, called the distorted shadow, the \textit{microlensed shadow}. In the case of relative motion between the source and the lens, the proposed distortion in the shadow will be a time-varying phenomenon.

In this work, we set constraints on PBH dark matter by exploiting recently studied phenomena of microlensing potentially distorting the black hole shadows imaged by Event Horizon Telescope (EHT) \cite{EHT_m871:2019dse, EHT_m872:2019uob, EHT_m873:2019jan, EHT_m874:2019ths, EHT_m875:2019pgp, EHT_m876:2019ggy, EHT_sgra1, EHT_sgra2, EHT_sgra3, EHT_sgra4, EHT_sgra5, EHT_sgra6}. Although the future space-based EHT-like telescope would be capable of observing such a distortion in the shadow of Sgr~A$^*$, a very small event rate (0.0014 per year) due to solar mass astrophysical compact objects will make it hard to observe such events~\cite{Verma:2023hes}. We constrain the possible enhanced population of lenses near the galactic center in the form of a PBH dark matter spike. 

Given that the intrinsic time-variability scale is large for M87 and assuming the typical time required to construct a photon ring is about 1 day, M87 provides an ideal case to search for time-variability of the photon ring due to PBH microlensing. In the absence of observing any asymmetry in the photon ring, a photon ring observed with an interferometry with 0.1 micro-arcsec resolution of the image would be sensitive to the PBH dark matter in the mass range $10^{-5}~M_\odot$ - $10^{6}~M_\odot$ with maximum sensitivity for $~10^3~M_\odot$ PBHs in the M87 galaxy, even if the PBH fraction of dark matter is as small as 0.01.

The remainder of this paper is organized as follows. Section~\ref{sec:MicroBHShadow} review and develops the theoretical framework for black hole shadow microlensing and deriving the microlened photon ring observable that quantifies photon ring distortions induced by passing PBHs. In section~\ref{sec:EventRate}, we introduce the lensing tube geometry in the context of microlensed shadow that defines the detection cross-section and calculate the expected microlensing event rate as functions of PBH mass, target source, accounting for both Galactic halo contribution and potential dark matter spikes around the SMBH. Section~\ref{sec:exclusion} presents our main results: conservative and optimistic constraints on the PBH dark matter fraction and PBH masses, demonstrating that next generation facilities can probe unprecedented parameter space and potentially detect individual microlensing events if PBH constitute more than 1\% of dark matter. Finally, section~\ref{sec:summary} summarizes the transformative potential of photon ring observations for probing PBH dark matter and outline future theoretical and observational work needed to realize this opportunity.

\section{Microlensing Black Hole Shadows}
\label{sec:MicroBHShadow}
In this section, we review the theoretical framework for gravitational microlensing of supermassive black hole (SMBH) shadows, building upon the formalism developed in our previous work.

\subsection{The shadow and the photon rings}
The shadow of a black hole represents one of the most direct observable signatures of the strong gravitational field of SMBH. It is crucial to distinguish between two related but distinct concepts: (1) \textit{critical curve (Bardeen shadow)}, the theoretical boundary defined by the envelope of light rays that asymptotically approach the photon sphere. This represents the shadow cast by the black hole's spacetime geometry when it is illuminated with background light sources uniformly. It only depends on the mass and spin angular momentum of the SMBH. (2) \textit{observational shadow}, The central brightness depression observed in the black hole images, which is influenced by both the underlying spacetime geometry and the emission properties of the surrounding accretion flow.

For a Schwarzschild black hole, the critical curve exhibits a characteristic circular shape of angular radius~\cite{2019PhRvD.100b4018G}
\begin{equation}
    R = 3\sqrt{3}\frac{G M}{c^2 D},
\end{equation}
where $M$ is the mass of the black hole, $G$ is the gravitational constant, and $D$ is the distance
from the observer to the black hole.
However, the shape can be asymmetric with the asymmetry reaching as large as $4\%$ for an extremally spinning black hole \cite{shadowkerr_Hioki:2009na}.

While the observed shadow may deviate from this idealized critical curve due to the accretion disk as a backlit, the critical curve remains observable as the inner boundary of photon ring structures. It is this photon ring boundary that provides the most robust probe of spacetime geometry and, consequently, the most sensitive target for microlensing studies.

The ring consists of an infinite series of increasingly demagnified images, with the brightest $n=1$ subring containing photons that have executed one-half orbit around the black hole before reaching the observer \cite{subrings}. Higher-order subrings ($n=2,3,...$) correspond to photons making additional orbits, each subring exponentially narrower and fainter than the previous one. Recent theoretical work has established that the ring width scales approximately as \cite{width_Gralla:2019xty}
\begin{equation}
    w_1\approx4.8~\mu{\rm as} \frac{M}{6.5\times10^9 M_\odot}\frac{16.8~{\rm Mpc}}{D_s}, 
\end{equation}
where $D_s$ is the distance from the observer to the SMBH 
for the $n=1$ subring. Thus, for M87 it turns out to be $4.8~\mu{\rm as}$, while for Sgr~A$^*$ the corresponding width is $1.4~\mu{\rm as}$. Resolving these subrings requires angular resolution better than their intrinsic width, presenting formidable technical challenges that drive the design of next-generation very long baseline interferometry (VLBI) facilities.
\subsection{Gravitational microlensing by foreground objects}
\begin{figure*}
    \centering
    \includegraphics[width=0.5\linewidth]{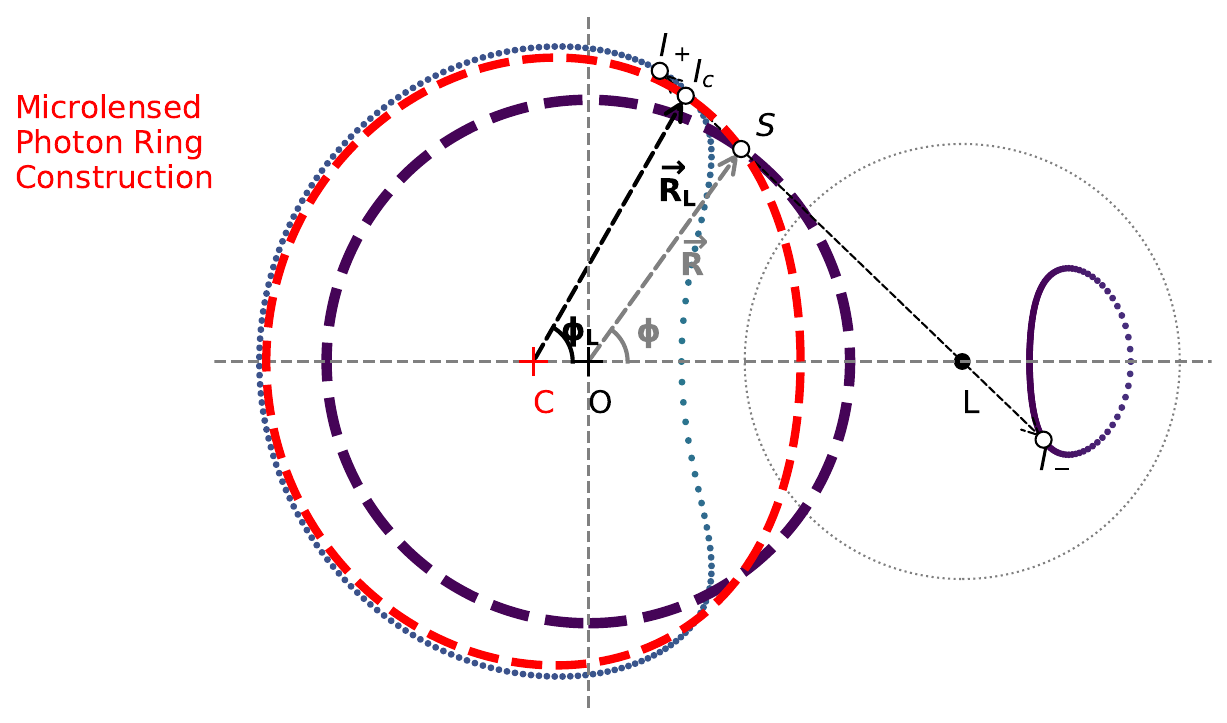}
    \includegraphics[width=1\linewidth]{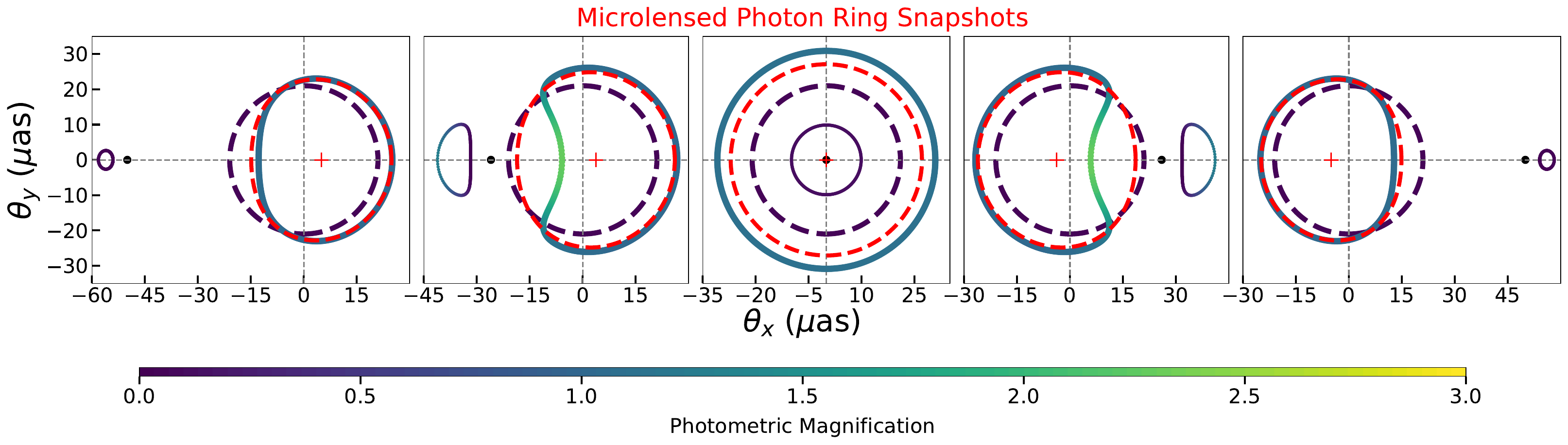}
    \caption{We demonstrate the phenomena of microlensing of the photon ring. As an illustration of the phenomena, we have taken the true boundary of the shadow of M87.}
    \label{fig:lensedshadowConstruction}
\end{figure*}

The presence of a compact object along the line of sight to a SMBH can gravitationally deflect light rays, resulting in apparent distortions of the observed shadow boundary. This microlensing effect has been proposed in ref.~\cite{Verma:2023hes} and forms the basis for our PBH detection methodology. The typical distortion arising due to microlensing is given by the angular Einstein angle,
\begin{equation}
\label{eq:thE}
    \theta_E = 2.5\textrm{~mas}\sqrt{\frac{M}{M_\odot} \frac{1~{\rm kpc}}{D_s}\left(\frac{D_s}{D_l} - 1\right)},
\end{equation}
where $M$ is the lens mass, $D_l$ is the distance to the lens, and $D_s$ is the distance to the source (SMBH). This encapsulates the fundamental scaling of microlensing effects with lens mass and geometry.

Under the weak lensing approximation, which is valid when the impact parameter of the light rays is much larger compared to the horizon of the black hole, the distorted shadow boundary can be calculated by treating each point on the original circular boundary as an individual source undergoing microlensing.

Consider an originally circular shadow of angular radius $R$ centered at the origin. A point on this boundary is parameterized as 
\begin{equation}
    \mathbf{R}(\phi) = \begin{bmatrix}
        R \cos{\phi} \\
        R \sin{\phi}
    \end{bmatrix},
\end{equation}
where $\phi$ is the azimuthal angle of a point source on the boundary of the shadow. The lens positioned at angular coordiante $\bm{\xi} = (\xi_x, \xi_y)$ relative to the shadow center. The source-lens separation for each boundary source point is $\bm{\beta} (\phi) = {\bm R}(\phi) - {\bm \xi}$ with magnitude $\beta(\phi) = \sqrt{\xi^2 + R^2 - 2 \xi R \cos(\phi)}$. Under gravitational lensing, each point on the original boundary is mapped to a new position given by the lens equation
\begin{eqnarray}
    \bm{\theta}_c(\phi) = \frac{\beta(\phi)^2 + 3\theta_E^2}{\beta(\phi)^2 + 2\theta_E^2}\bm{\beta}(\phi).
\end{eqnarray}
The image position relative to the original coordinate system is
\begin{eqnarray}
\bm{OI}_c(\phi) &=& \bm{\theta_c} + \bm{\xi}= \bm{R} + \frac{\theta_E^2}{\beta(\phi)^2 + 2\theta_E^2}\bm{\beta}(\phi).
\end{eqnarray}

The microlensing not only distorts the shadow shape but also shifts its apparent center. The new shadow center can be calculated as 
\begin{eqnarray}
	\label{eq:lcenter}
    \bm{OC} &\equiv& \frac{\bm{OI}_c(\pi) + \bm{OI}_c(0)}{2}.
\end{eqnarray}
This definition captures the centroid shift by averaging the positions of opposite points on the distorted boundary. The final distorted shadow boundary, referenced to this new center, is 
\begin{eqnarray}
\label{eq:RL}
\bm{R}_L = \bm{OI}_c - \bm{OC}= \bm{R}- \bm{OC} + \frac{\theta_E^2}{\beta(\phi)^2 + 2\theta_E^2}\bm{\beta}(\phi).
\end{eqnarray}
The angular position of each point on the distorted boundary is
\begin{equation}
\label{eq:phiL}
\phi_L = \tan^{-1}\left(\frac{R_{L,y}}{R_{L,x}}\right).
\end{equation}

\subsection{Observable Signatures}
There are three key observable effects of the microlensing-induced distortions in the black hole shadow.
\subsubsection{Shift in the center}
\begin{eqnarray}
\label{eq:lcenterdetailed}
\bm{OC} &=& -\frac{\theta_E^2(2\theta_E^2 - R^2 + \xi^2)}{[2\theta_E^2 +(R - \xi)^2][2\theta_E^2 +(R + \xi)^2]}\bm{\xi}.
\end{eqnarray}

\subsubsection{Magnification in size}
We define the radius of the microlensed shadow as the average of $R_L(\phi)$ over $\phi$. Hence, the average radius is given by
\begin{eqnarray}
\langle R_L \rangle&=& \frac{1}{2\pi}\int_0^{2 \pi} |\bm{R}_L(\phi)|d\phi.
\end{eqnarray}
\subsubsection{Asymmetry in shape}
Unlike the shift in the center and the magnification in the size of the shadow, asymmetry is an absolute observable effect of the microlensing of the shadow. Hence, it can be measured even in a single epoch frame of the shadow. The asymmetry of the lensed shadow is defined as the variance in the observed radius of the shadow and given by,
\begin{eqnarray}
	A &=& \sqrt{\frac{1}{2\pi}\int_0^{2\pi} (R_L(\phi)^2 - \langle R_L \rangle^2)d\phi}.
\end{eqnarray}
\begin{figure}
    \centering
    \includegraphics[width=\linewidth]{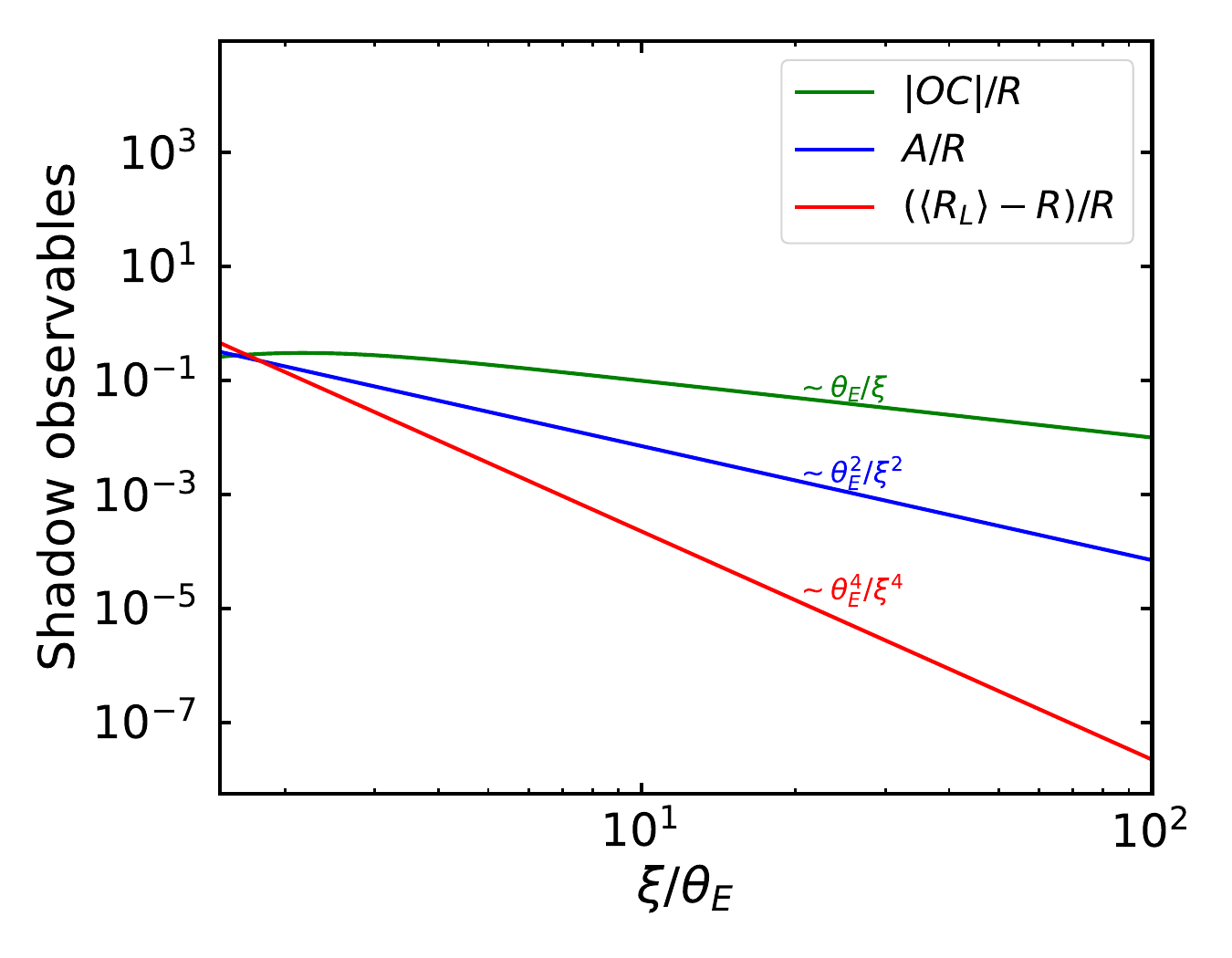}
    \caption{Shadow observables as a function of the lens impact parameter $\xi/theta_E$. Plotted are the normalized centroid displacement \(\lvert O C\rvert / R\), the normalized asymmetry \(A/R\), 
    and the relative change in radius \(( \langle R_L \rangle - R)/R\) as the lens approaches the line of sight. 
    }
    \label{fig:shadobs}
\end{figure}
  In \cref{fig:shadobs}, we illustrate how the above three key shadow observables vary with the normalized impact parameter $\xi/\theta_E$ when a compact object with Einstein angle~$\theta_E$ gravitationally lenses a black hole shadow of radius~$R$. Each observable exhibits a power-law dependence on the impact parameter. The center-shift declines as $\theta_E/\xi$, the asymmetry falls as $\theta_E^2/\xi^2$, and the radius of the shadow change decreases as $\theta_E^4/\xi^4$ (the steepest decay), as indicated by the color-coded annotations on the right of the plot. As seen from the figure, at a given impact parameter $\xi/\theta_E$, asymmetry of the lensed shadow can provide a better observational significance than the magnification of the size, because of its higher lensing cross-section. Therefore, in this work, we use asymmetry as the shadow observable to study the microlensing event rate due to PBHs.


\section{Microlensing event rate calculation}
\label{sec:EventRate}
The detectability of PBH dark matter through black hole shadow microlensing depends on the expected event rates. In this section, we describe the calculation of the microlensing event rate for the shadows of SgrA~$^*$ and M87 due to PBH dark matter distribution towards them. The key question we address is given an interferometric facility with angular resolution characterized by $\sigma_A/R$, observing for time $T_{\rm obs}$ with cadence $t_{\rm cad}$ and exposure time $t_{\rm exp}$, how many microlensing events should we expect for a PBH population with mass $M_{\rm pbh}$ comprising fraction $f_{\rm pbh}$ of dark matter?

\subsection{Detection criteria}
\label{subsec:DetCriteria}
The detectability of microlensing signatures in a BH shadow is governed by the induced distortion and the precision with which we can measure the distortion. Microlensing by a foreground lens produces three primary observable effects: asymmetry in the shadow shape, magnification of the shadow size, and displacement of the center of the shadow. Among these observables, the asymmetry signal is the most sensitive due to its favorable scaling with impact parameter.

For a lens at angular separation $\xi$ from the shadow center, the induced asymmetry scales approximately as 
\begin{eqnarray}
    A  &\approx& \frac{R}{\sqrt{2}}\frac{\theta_E^2}{\xi^2},
\end{eqnarray}
where $R$ is the unperturbed/unlensed shadow radius. This inverse-square dependence on the impact parameter means that asymmetry remains detectable even for relatively distant lens parallaxes. In contrast, the size magnification exhibits a steeper falloff, as given by
\begin{eqnarray}
    \langle R_L \rangle &\approx& R \left(1+ \frac{9}{4}\frac{\theta_E^4}{\xi^4}\right),
\end{eqnarray}
making it sensitive only to very close encounters. 

We define the detection threshold by requiring the induced asymmetry to exceed the measurement uncertainty characterized by $\sigma_A$. Setting $A(\xi_{\rm th}) = \sigma_A$ and solving for the maximum impact parameter that yields detectable asymmetry, we obtain,
\begin{eqnarray}
\label{eq:xith}
\xi_{\textrm{th}} &=& 
    26.6 \theta_E \left(\frac{10^{-3}}{\sigma_a/R}\right)^{1/2},
\end{eqnarray}

This threshold defines a circular detection region around the shadow center with radius $\xi_{\rm th}$. Any lens passing within this radius during an observation will induce observable asymmetry. The corresponding threshold for size magnification is $0.7 \theta_E \left( \frac{10^{-3}}{\sigma_a/R}\right)^{1/4}$, which is smaller by a factor of approximately $38\left(10^{-3}/(\sigma_a/R)\right)^{1/4}$. Since the detection cross-section scales as $\sigma_{\rm det} \propto \xi_{\rm th}^2$, asymmetry measurements provide about 1400$\times$ larger effective cross-section than size magnification measurements for an EHT-like telescope precision $\sigma_{\rm A}/R=10^{-3}$.

Consequently, we adopt the asymmetry observable for the primary detection criteria throughout this analysis. The subset of events producing detectable size magnification represents a rare population of close encounters that comprise only approximately $0.07$~\% of the total event rate for $\sigma_A/R=10^{-3}$. However, these rare close encounters, when detected, would further facilitate unambiguous detection of a microlensing signature, which will be difficult to mimic through other astrophysical processes.

An additional temporal constraint arises from the requirement that microlensing events be resolvable in time. The Einstein crossing time $t_E=\theta_E D_l/v_\perp$, where $v_\perp$ is the transverse velocity of the lens, must be compared with the observational cadence $t_{\rm cad}$ (frequency of taking different frames of BH images). Events with $t_E<<t_{\rm cad}$ will appear as instantaneous shape changes, which might fall between the two epochs of observation, while events with $t_E>>t_{\rm cad}$ can show a movie of the microlensing signatures and distinguished from intrinsic source variability throughout the characteristic evolution. The interplay between event duration, observing cadence, exposure time, and intrinsic variability timescales will create a complicated selection function that we address in detail below. However, we will assume no intrinsic source variability.


\subsection{PBH lens population}
\label{sec:lenspop}

We focus on already observed shadows by EHT, i.e., SgrA and M87. While calculating the event rates for these two black holes, we consider the dark matter distribution in the form of a standard NFW profile and a dark matter spike near the black hole. For a more realistic picture, we consider the possible co-existence of the two dark matter profiles in the foreground. 
For example, for SgrA, we can have two types of populations of PBHs as dark matter.  The first is the NFW contribution 
and the second is the PBH spike forming around the central SMBH. On the other hand, for M87, the first and second contributions come from the NFW and the spike within the M87 galaxy, but there is the third one coming 
from the Milky Way PBH dark matter in the foreground of the M87 galaxy. 
We first briefly discuss the profiles of the DM spike and the NFW.

\subsubsection{Dark matter spike profile}
Let us consider an initial density profile of DM to be,
\begin{eqnarray}
    \rho_\textrm{DM} &=& \rho_0 \left(\frac{r_0}{r}\right)^\gamma,
\end{eqnarray}
where $\gamma$ is the power law index and $\rho_0$ and $r_0$ are the DM halo parameters. We can then obtain a DM spike formed adiabatically around the SMBH of mass $M_\textrm{BH}$ with a density profile~\cite{Gondolo:1999ef},
\begin{eqnarray}
    \rho_\textrm{DM}^\textrm{sp} &\approx& \rho_\textrm{sp} \left(\frac{R_\textrm{sp}}{r}\right)^{\gamma_\textrm{sp}} = \rho_b\left(\frac{r_b}{r}\right)^{\gamma_\textrm{sp}},
\end{eqnarray}
where 
\begin{eqnarray}
    \rho_\textrm{sp} &=& \rho_0\left(\frac{R_\textrm
    {sp}}{r_0}\right)^{-\gamma},\nonumber \\
    R_\textrm{sp} &=& \alpha_\gamma r_0\left(\frac{M_\textrm{BH}}{\rho_0 r_0^3}\right)^\frac{1}{3-\gamma}, \nonumber \\
    \gamma_\textrm{sp} &=& \frac{9 - 2\gamma}{4 - \gamma}, \nonumber
\end{eqnarray}
where the parameters of the density spike $\rho_\textrm{sp}$ and $R_\textrm{sp}$ are the density and radius of the spike, respectively, at the outer edge. Instead of $\rho_\textrm{sp}$ and $r_\textrm{sp}$, one can use $\rho_b$ and $r_b$, which are the density and radius of the spike at its inner edge. It has been shown in ref.~\cite{Daghigh:2022pcr} that the $\alpha_\gamma$ can be related to the spike parameters as
\begin{equation}
    \alpha_\gamma = \left(\frac{\rho_\textrm{sp} R_\textrm{sp}^3}{M_\textrm{BH}}\right)^\frac{1}{3-\gamma}.
\end{equation}
\subsubsection{NFW profile}
DM density is given by a standard spherically-symmetric NFW profile~\cite{NFWparameter} about the Galactic center, 
\begin{equation}
\rho_{\textrm{DM}} = \frac{\rho_0}{\frac{r}{r_0}\left(1+ \frac{r}{r_0}\right)^2},
\end{equation}
where $\rho_0 =1.06\times10^7 \textrm{ M}_{\odot}/\textrm{kpc}^3 $ and $r_0 = 12.53 \textrm{ kpc}$ is the DM scale radius. The distance $r$ of a lens from the center of the Galaxy can be written in terms of the distance of the lens from the earth's position $D_l$, as $r = D_s - D_l$.

\subsection{The optical depth and Event rate}
\begin{figure*}
\centering
\includegraphics[width=\textwidth]{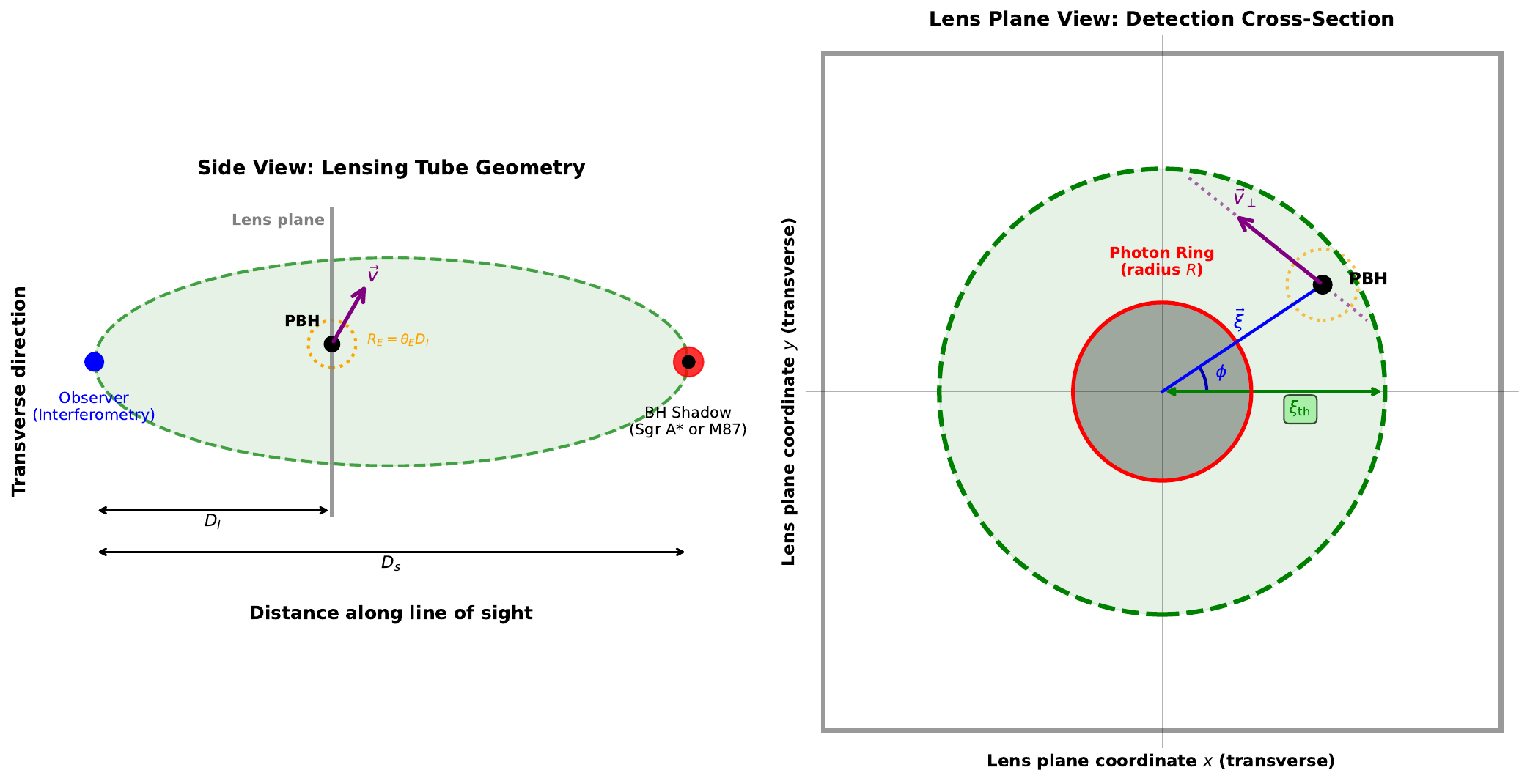}
\caption{Geometric configuration for microlensing of black hole shadows. \textit{Left panel:} Side view showing the lensing tube extending from the observer to the black hole shadow at distance $D_s$. The tube has cross-sectional radius $R_{\rm th} = \xi_{\rm th} D_l$ at lens distance $D_l$, defining the region within which a PBH lens can induce detectable shadow distortions. A PBH moving with transverse velocity $\vec{v}_{\perp}$ will cause microlensing when its trajectory intersects the tube. \textit{Right panel:} View in the lens plane showing the detection cross-section. The photon ring (red circle, radius $R$) marks the shadow boundary. Any PBH with impact parameter $|\vec{\xi}| < \xi_{\rm th}$ (within the green dashed circle) will induce asymmetry exceeding the detection threshold $\sigma_A$. The PBH trajectory (purple dotted line) shows the lens motion through the sensitive region with velocity $\vec{v}_{l\perp}$.}
\label{fig:MLrateGeometry}
\end{figure*}

The calculation of expected microlensing event rates requires careful consideration of the geometric configuration connecting the observer (interferometry), the foreground lens population, and the background black hole shadow as shown in fig.~\ref{fig:MLrateGeometry}.
Unlike traditional stellar microlensing, where point sources are magnified, our scenario involves the distortion of an extended photon ring boundary. We can conceptualize this geometry through the notion of a ``\textit{lensing tube}''--a cylindrical region extending along the line of sight to the black hole shadow, within which any passing PBH will induce detectable asymmetry in the shape of the photon ring/shadow. The cross-sectional radius of this tube at distance $D_l$ from the observer is given by the physical threshold distance $R_{\rm th} = \xi_{\rm th} D_l$, where $\xi_{\rm th}$ is the angular threshold derived in Section~\ref{subsec:DetCriteria}. A PBH entering this tube at any point along the line of sight will have its trajectory bring it within the critical impact parameter of the shadow boundary, thereby inducing observable distortions. The probability that a lens lies within this tube at any given instant defines the \textit{optical depth}, while the rate at which lenses enter and traverse the tube determines the \textit{event rate}.

We begin by establishing the optical depth, which quantifies the instantaneous probability that at least one PBH from the foreground population lies within the lensing tube. Consider a black hole shadow at distance $D_s$ from the observer, which we take to be either Sgr A$^*$ at $D_s=8.2$~kpc or M87 at $D_s=16.8$~Mpc. The lensing tube extends from the observer position to the source with cross-sectional area $\pi R_{\rm th}^2$ at lens distance $D_l$. The volume element of the tube between distances $D_l$ and $D_l+dD_l$ is $dV=\pi \xi_{\rm th}^2D_l^2dD_l$. If PBHs comprise a fraction $f_{\rm pbh}$ of the dark matter with spatial density $\rho_{\rm DM}(D_l,l,b)$ and individual mass $M_{\rm pbh}$, 
then the number density of PBH lenses is $n_{\rm pbh} = f_{\rm pbh}\rho_{\rm DM}/M_{\rm pbh}$. This number density can be taken constant throughout a cross-sectional area of the tube at a distance $D_l$, since $R_{\rm th}$ will remain very small compared to the scale of DM density variation. The expected number of PBHs in the volume element is $dN=n_{\rm pbh}dV$, the optical depth follows from integrating along the entire line of sight: $\tau=\int_0^{D_s}n_{\rm pbh}(D_l,l,b) dV$. Substituting the asymmetry threshold angular radius $\xi_{\rm th}$, given by eq.~\ref{eq:xith}, and the angular Einstein radius (eq.~\ref{eq:thE}, we obtain an optical depth that depends on both the PBH mass and the astrometric precision as
\begin{eqnarray}
\label{eq:tau}
    \tau&=&\int_0^{D_s}\frac{f_{\rm pbh}}{M}\rho_{\rm DM}(D_l,l,b)\pi \xi_{\rm th}^2 D_l^2dD_l
\end{eqnarray}

Similar to traditional photometric microlensing of stellar sources, the optical depth is independent of lens mass. The optical depth also scales inversely with the BH image resolution, such that improving from $\sigma_A/R=10^{-3}$ to $10^{-4}$ increases the optical depth by a factor of ten, corresponding to the increased effective tube cross-section accessible to more precise measurements.

For our two primary targets, the optical depth integral in eq.~\ref{eq:tau} receives contributions from distinct dark matter populations. For Sgr~$A^*$, we integrate the Milky Way NFW profile $\rho_{\rm MW}(D_l)$ from $D_l=0$ to $D_s=8.2$~kpc, where the distance from the Galactic center is $r=D_s-D_l$, being the Sun's Galactocentric radius and $(l,b)=(0,0)$ the Galactic coordinates towards Sgr~A$^*$. Additionally, if a dark matter spike exists around Sgr~A$^*$, we include a separate contribution by integrating the spike profile from $r_{\rm min}=4 R_s$ to $r_{\rm max} = R_{\rm sp}$. For M87, the situation becomes more complex with three dominant contributing populations. First and second, the M87 spike and NFW halo contributions are computed similarly to Sgr~A$^*$ but at the much larger source distance $D_s=16.8$~Mpc. Third, and crucially different from Sgr~A$^*$, the Milky Way NFW halo provides a foreground contribution by integrating from $D_l=0$ to approximately 50~kpc, beyond which the Milky Way halo density becomes negligible. There will also be a fourth negligible contribution of intergalactic dark matter between the Milky Way and the M87 galaxy, which we will ignore in our calculation.

The NFW and spike contribution is given by:
\begin{equation}
    \tau_{\rm sp} = \frac{8.9  G}{c^2} f_{pbh} \frac{R}{\sigma_a} \rho_{0,\rm sp}\left(\frac{r_{0,\rm sp}}{D_s}\right)^{\gamma_{\rm sp}} D_s^2\frac{1}{6-5\gamma_{\rm sp}+\gamma_{\rm sp}^2}
\end{equation}

\begin{equation}
    \tau_{\rm NFW} = \frac{8.9  G}{c^2} f_{pbh} \frac{R}{\sigma_a} \rho_0\left(1+\frac{r_0}{D_s}\log{\frac{r_0}{D_s+r_0}}\right)r_0^2
\end{equation}

The optical depth quantifies the instantaneous probability of finding a PBH within the lensing tube, but to calculate the event rate, we must account for the dynamics of lenses crossing through the tube over time. A PBH moving with transverse velocity $v_\perp$ relative to the the lensing tube will traverse the a segment of the tube of length $L=2D_l\xi_{\rm th}\cos{\alpha}$ in a crossing time $t_{\rm event}=L/v_\perp$, where $\alpha$ is the angle at which the the projected lens enters the sensitive region on a lens plane. The rate at which new lenses enter the tube is determined by the flux of lenses $n_{\rm pbh}v_\perp$ multiplied by the tube cross-sectional area. Following the formalism of Griest (1991)~\cite{1991ApJ...366..412G} (which has been a standard formalism for microlensing rate calculation),           
we construct the differential event rate by considering the geometric configuration shown in fig.~\ref{fig:MLrateGeometry}. A PBH at distance $D_l$ from the observer, moving with velocity ${\bf v_l}$ in the lens plane (defined perpendicular to the line of sight), will cause a detectable microlensing event if its trajectory brings it within impact parameter $R_{\rm th}$ of the shadow center. We parameterize the lens velocity relative to the lensing tube by its magnitude $v_{l \perp}$ (the component perpendicular to the line of sight) and the angle $\alpha$ at which the lens enters the sensitive region. The velocity distribution of PBH lenses depends on their orbital dynamics within the dark matter halo, which we model as an isotropic Maxwellian distribution with velocity dispersion $v_c(r)$ appropriate for the local dark matter density at center radius $r$ from either the Galactic center or the M87 center.

The differential event rate can be expressed as:
\begin{align}\label{eq:EventRate}
    d\Gamma =& 2\pi f_{\rm pbh} \rho(D_l,l,b)\frac{\psi(M)}{M} R_{\rm th} \cos{\theta} v_{l\perp}^2 f_\perp(v_{l\perp}) \\\nonumber
    &dD_l dM d\theta dv_{l\perp},
\end{align}
where $f_\perp(v_{l\perp})$ is the probability distribution for the transverse velocity component. The factor $2\pi R_{\rm th}dD_l$ is the surface element of a cylindrical shell at distance $D_l$ with thickness $dD_l$, the factor $\cos{\phi}$ projects the lens velocity onto the direction perpendicular to the tube boundary, and $v_{l\perp}^2$ enters because the flux of particles is proportional to velocity and we integrate over velocity space with measure $v_{l\perp}dv_{l\perp}$ in cylindrical coordinates. For an isotropic Maxwellian velocity distribution, $f_\perp(v_{l\perp}) = (2/(\pi v_c^2))exp(-v_{l\perp}^2/v_c^2)$ where $v_c(r)$ is the circular velocity at the lens position, which we can compute from the dark matter density profile as $v_c(r) = \sqrt{G M_{\rm DM}(<r)/r}$ where $M_{\rm DM}(<r)$ is the enclosed dark matter mass within radius $r$.

To connect the event rate to observable quantities, we relate the geometry to the Einstein crossing time $t_E= R_E/v_{l\perp}$. The angle $\phi$ parameterizes the impact parameter $u_0=u_T\sin{\phi}$, such that lenses entering with $|\phi|<\pi/2$ will pass within the detection threshold. By chnaging variable from $\phi$ to the minimum impact parameter $u_0$, we obtain
\begin{align}
\frac{d\Gamma}{dt_E} &=& 4\pi f_{\rm PBH} \int_0^{D_s} dD_l \int_0^{u_T} \frac{du_{\rm min}}{\sqrt{u_T^2 - u_{\rm min}^2}} \frac{\rho_{\rm DM}(D_l)}{M_L v_c^2(r[D_l])}\nonumber\\
&&v_{l\perp}^4 \exp\left(-\frac{v_{l\perp}^2}{v_c^2}\right)
\label{eq:dgamma_dte_umin}
\end{align}
where now $v_{l\perp}=2R_E\sqrt{u_T^2-u_0^2}/t_E$. The total observable event rate over an observing campaign of duration $T_{\rm obs}$ is obtained by integrating over the relevant range of Einstein crossing times and multiplied by the observing time: $N_{\rm events} = T_{\rm obs} \int_{t_E,\rm min}^{t_E,\rm max}(d\Gamma/dt_E)dt_E$, where the integration limits are determined by the temporal resolution capabilities of the observing campaign (events much shorter than the cadence/exposure-time become undetectable) and the total campaign duration (events much longer than $T_{\rm obs}$ have low probability of being captured).

For the velocity dispersion $v_c(r)$ entering these expressions, we adopt different prescriptions depending on whether the lens resides in the spike region or the extended NFW halo. In the spike region around either Sgr~A$^*$ or M87, lenses follow approximately Keplerian orbits with $v_c(r) = \sqrt{GM_{\rm SMBH}/r}$, yielding $v_c \approx 600$~km/s at $r=1$~pc for Sgr~A$^*$ and $v_c \approx 6000$~km/s at 10~pc for $M87$. These high velocities in the spike region lead to short Einstein times, creating a population of rapid events that may be difficult to distinguish from intrinsic source variability. In the extended NFW halo region, we adopt constant characteristic velocities: $v_c=220$~km/s for the Milky Way halo and $v_c=500$~km/s for the M87 halo, based on their respective rotation curves and velocity dispersion measurements~\cite{sgravc_McMillan_2016, M87vc_Cohen:1999ii}. These velocities remain approximately constant throughout the halo, producing a population of intermediate-duration events more amenable to detection with day-scale observing cadences.

\subsection{Event rate for SgrA and M87}
The event rate calculation must be performed separately for each dark matter component and summed to obtain the total rate. For Sgr~A$^*$, we compute $d\Gamma_{\rm total}/dt_E = d\Gamma_{\rm spike}/dt_E+d\Gamma_{\rm NFW, MW}/dt_E$, where the spike contribution integrates from $r_{\rm min}=4R_s$ to $R_{\rm sp}$ and the Milky Way contribution integrates from the observer to $D_s-R_{\rm sp}$. For M87, the total rate is $d\Gamma_{\rm total}/dt_E= d\Gamma_{\rm spike}/dt_E + d\Gamma_{\rm NFW, M87}/dt_E + d\Gamma_{\rm NFW, MW}/dt_E$, with the M87 contribution intergrating over the M87 galaxy and the Milky Way contribution providing the cruicial foreground channel.

The relative importance of these components varies dramatically with PBH mass and the resolution of the interferometry. For solar-mass PBHs observed with the next generation precision $\sigma_A/R\sim 10^{-4}$, the spike contribution dominates by factors of 10-100 if a steep (with $\gamma_{\rm sp}\simeq 2.23$) spike exists, but the NFW contribution provides robust lower limits independent of uncertain spike physics. For M87 observations, the Milky Way foreground can contribute comparably to or even exceed the M87 spike rate for $M_{\rm PBH} <10^{-4}~M_\odot$, despite the $\sim 2000$ times greater distance, because the foreground integration path samples a substantial column of Milky Way dark matter.

\begin{figure*}
	\centering
	\includegraphics[width=0.45\linewidth]{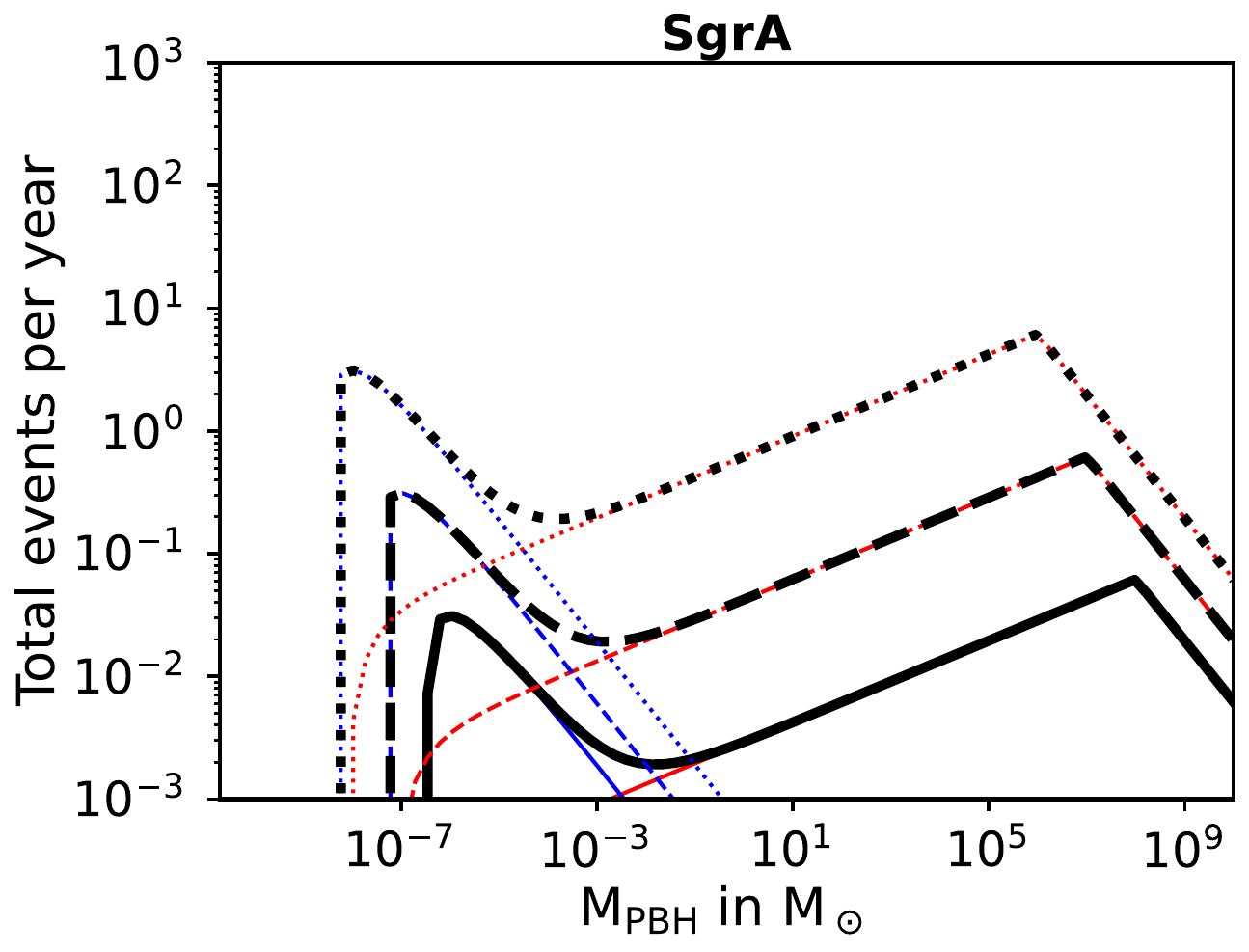}
	\includegraphics[width=0.45\linewidth]{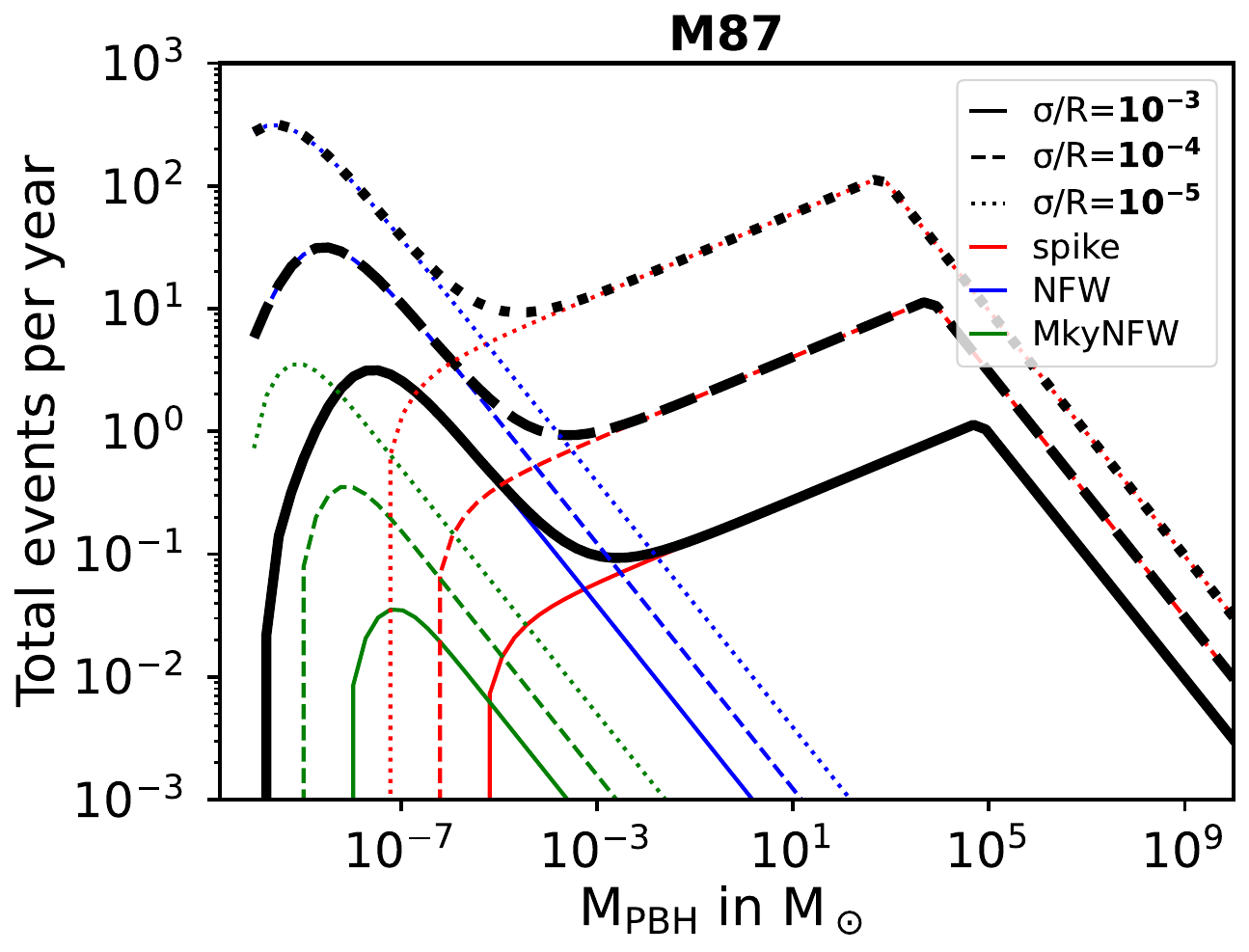}    
	\caption{Predicted microlensing event rates for PBHs near SgrA* (left) and M87* (right) due to PBH lenses in the foreground. Different dark matter density profiles shown in different colors: density spike (red), standard NFW (blue), and MilkyWay NFW (green) in the case of M87. The thick, solid black curve in each panel represents the envelope of combined density profiles. Different line styles show different possible telecsope resolution $\sigma/R$. The spike profile yields the highest event rates, reaching values of order $10^{1}$--$10^{2}$ events per year for $M_{\rm PBH}\sim 10^{3}$--$10^{5}\,M_\odot$ in the case of M87}. 
    \label{fig:EventRateM}
\end{figure*}
    
Figure~\ref{fig:EventRateM} presents our calculated event rates as a function of PBH mass for various levels of 
precision of measuring the asymmetry of the photon ring and dark matter contributions. The event rate peaks at a characteristic PBH mass for each observational configuration, determined by the balance between detection cross-section (favoring heavier masses with larger $\theta_E$) and velocity-weighted number density (favoring lighter masses with higher $n_{\rm pbh}$). For EHT-like precision such that $\sigma_A/R\sim 10^{-3}$, M87 observations achieve maximum sensitivity near $10^5~M_\odot$ with event rate approaching $\sim 0.01-0.1$ per year for percent-level PBH fractions in optimistic spike scenarios. The future EHT-like facilities with asymmetry uncertainty improved by ten-fold shift the sensitivity peak to $M_L\sim 10^{4}~M_\odot$ while increasing rates by an order of magnitude. Further improvements to $\sigma_A/R$ could probe 100~$M_\odot$ PBH with event rates potentially exceeding unity per year, enabling statistical studies of the PBH mass function through the distribution of observed Einstein angle and Einstein times. The stark difference in event rates between Sgr~A$^*$ and M87 (with M87 yielding 10-100 times higher rates for comparable precision) primarily reflects the longer intrinsic variability time scales of M87 (months versus hours, which enable detection across a much broader range of event durations despite the disadvantage of a larger distance. This temporal consideration, rather than simple optical depth arguments, ultimately determined the relative power of the two targets for constraining PBH dark matter through searching for a microlensing signature in the movie of photon rings upcoming in the near future.

\begin{figure*}
	\centering
    \includegraphics[width=0.45\linewidth]{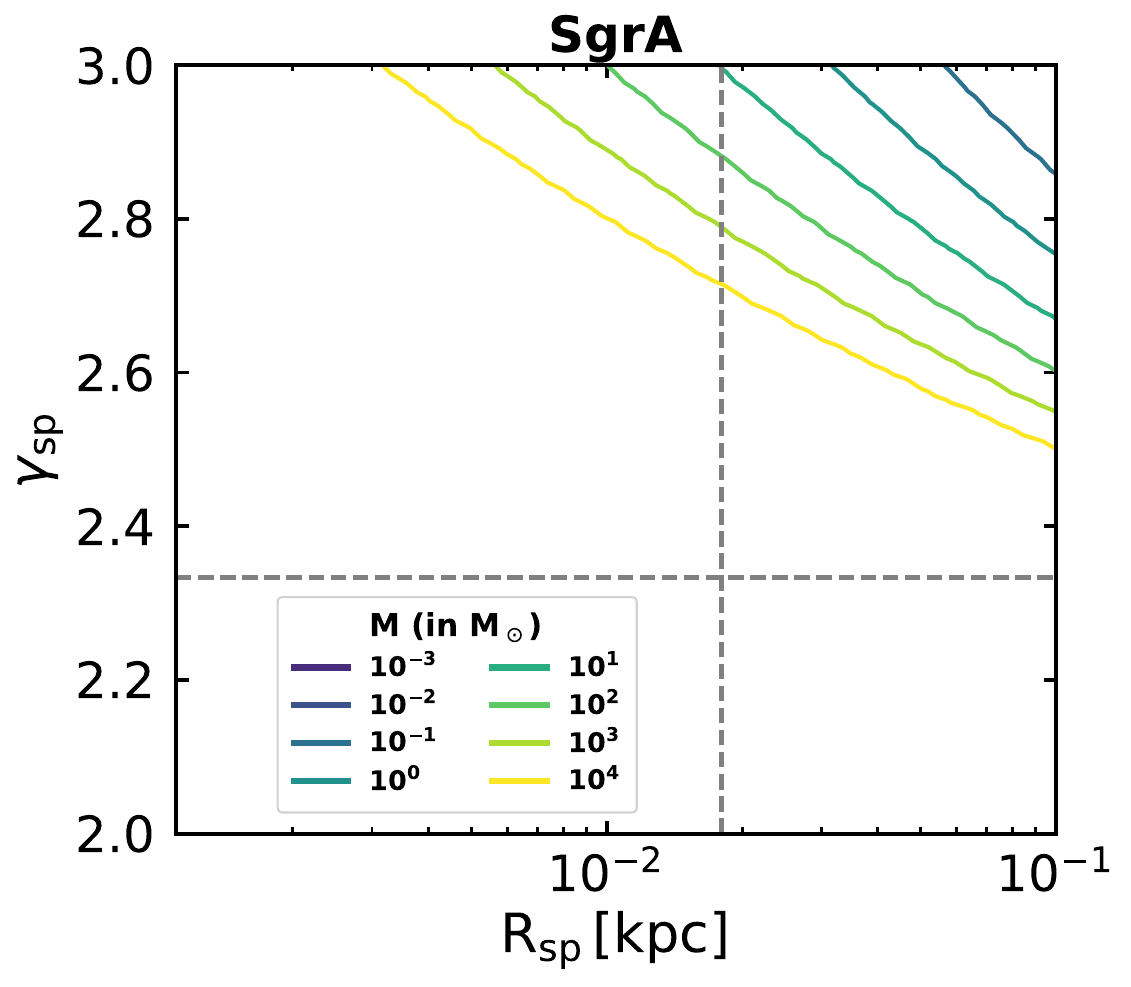}
	\includegraphics[width=0.45\linewidth]{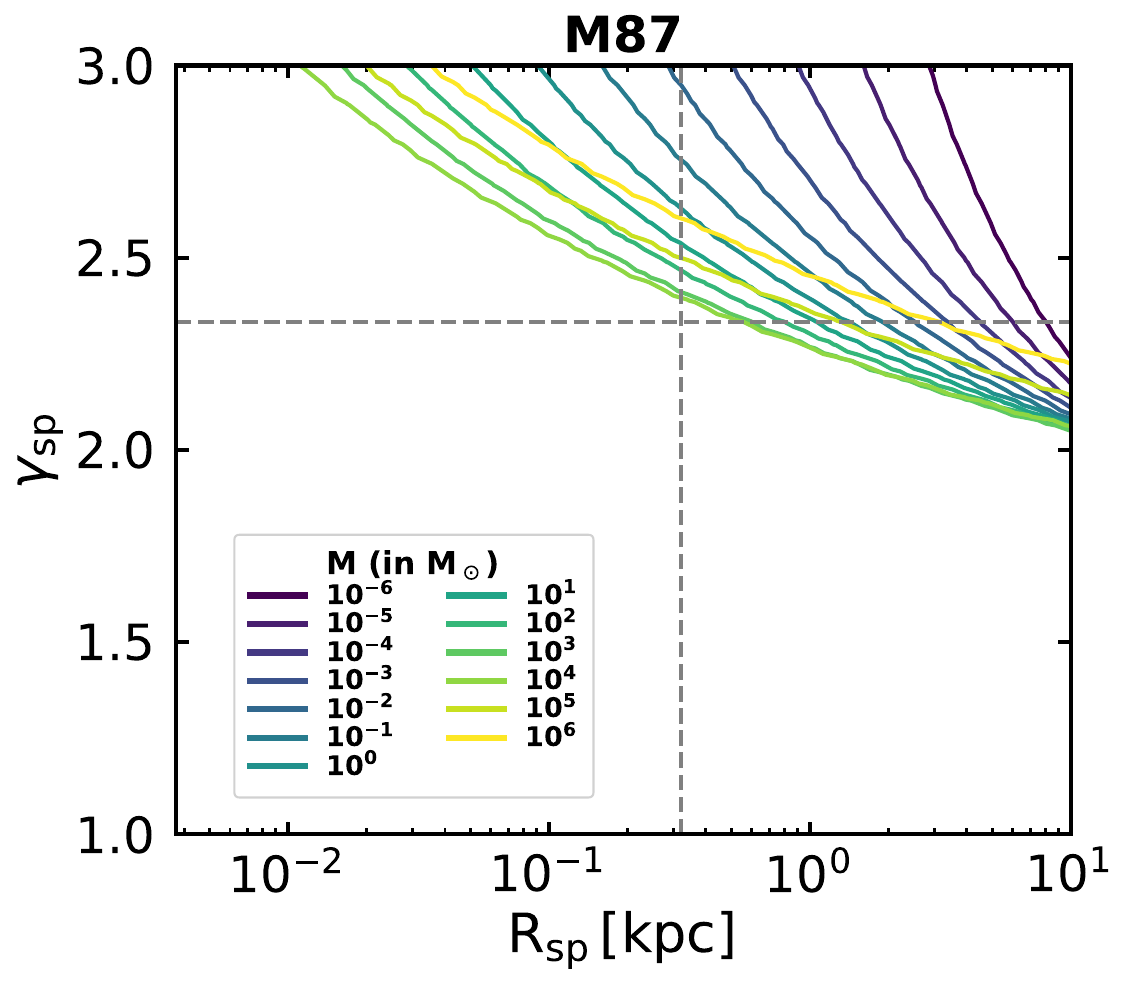}
	\caption{\textit{Right panel:} Parameter space for the SgrA dark‐matter spike at fixed $\sigma_A / R = 10^{-4}$ that gives rise to the total number of microlensing events per year $>1$. The horizontal axis shows the spike radius $R_{\rm sp}$ in kiloparsecs, and the vertical axis shows the inner density slope $\gamma_{\rm sp}$. Iso‐mass contours for primordial black holes are overlaid, each labeled by the PBH mass ranging from $10^{-4}$-$10^{4}$ $M_\odot$. The range of $R_{\rm sp}$ is determined from the fact that the mass enclosed in the volume of spike radius. Similarly, \textit{right panel:} shows the parameter space for the M87 dark‐matter spike. Iso‐mass contours for primordial black holes are overlaid, each labeled by the PBH mass ranging from $10^{-6}$-$10^{6}$ $M_\odot$. The dashed horizontal line marks the slope $\gamma_{\rm sp}\simeq 7/3$ expected from adiabatic SMBH growth, while the dashed vertical line indicates the maximum allowed spike radius set by the requirement that the enclosed spike mass remains  consistent with the host halo.}
    \label{fig:DifferentialEvents}
\end{figure*}

We calculate the total event rate using \cref{eq:EventRate}, assuming a monochromatic distribution of PBHs, for a given telescope resolution $\sigma/R$. 
\begin{equation}
N_{\text{tot}}(M, \sigma/R) = \int_{r_{\min}}^{r_{\max}} \frac{d\Gamma}{dr}(r, M, \sigma/R) \, dr
\end{equation}
The integration variable $r=D_s -D_l$ is the distance from SMBH to PBH lens. We consider different possible dark matter contributions to the lensing for SgrA and M87. For example, for M87, dark matter contribution will be a sum of the contributions from the DM spike near its center and the NFW profiles in the containing galaxy, as well as in the observer's galaxy, i.e., the  Milky Way galaxy. On the other hand, for SgrA, it is just the sum of the DM spike near its center and the NFW profile in the Milky Way galaxy. Hence, the integration limits $r_{\text{min}}$, $r_{\text{max}}$ will be different for each BH depending upon the DM contributions.


In \cref{fig:DifferentialEvents}, we show the expected microlensing events as a function of PBH mass for two supermassive black hole shadows: Sgr A* (Milky Way center) and M87* (center of M87 galaxy). The figure depicts how often the PBHs forming a fraction of dark matter $f_\textrm{PBH}$ will cross in front of the shadow and detectably distort it each year. We show the rate for different DM profiles and telescope resolution $\sigma/R$.

We choose to show the events that last longer, i.e, with $ t_e > 1 \text{ day}$, to be in the ballpark of 
observation cadence and image reconstruction time of the telescopes. For example,  EHT observes for $\sim 1$ week per year, not continuously, and takes $\sim$ hours to produce a shadow image. The microlensing event rate produced by PBHs depends sensitively on three main physical inputs: the PBH mass $M_{\rm PBH}$, the distribution of dark matter around 
the central supermassive black hole (SMBH), and $\sigma/R$. 
The lensing cross-section therefore grows linearly with mass:
\begin{equation}
    \sigma_{\rm lens} \propto \theta_E^2 \propto M_{\rm PBH}.
\end{equation}
with  $\theta_E \propto \sqrt{M_{\rm PBH}}$.  At the same time, the number density of PBHs decreases for larger masses,
\begin{equation}
    n_{\rm PBH} = \frac{f_{\rm PBH}\,\rho_{\rm DM}}{M_{\rm PBH}} \propto M_{\rm PBH}^{-1}.
\end{equation}
Thus, the event rate scales approximately as
\begin{equation}
    \Gamma \,\propto\, n_{\rm PBH} \, \theta_E \,\propto\, 
    \frac{1}{M_{\rm PBH}}\, \sqrt{M_{\rm PBH}} 
    = M_{\rm PBH}^{-1/2}.
\end{equation}
Consequently, the event rate initially rises with $M_{\rm PBH}$ due to the increasing
lensing area, but eventually falls as the PBHs become too rare. This produces a characteristic peak in $\Gamma(M_{\rm PBH})$, as seen in our results. The microlensing rate depends critically on the spatial number density of PBHs, which in turn follows the underlying dark matter density profile near SgrA*. Since the PBH number density scales as
\[
n_{\rm PBH}(r) = \frac{f_{\rm PBH}\,\rho_{\rm DM}(r)}{M_{\rm PBH}}\,.
\]
Any enhancement or suppression of \(\rho_{\rm DM}(r)\) near the Galactic Center has a direct and often dramatic impact on the event rate. For this reason, the plot compares several distinct dark matter profiles: the standard NFW profile behaves as \(\rho\propto r^{-1}\) in the inner region and \(\rho\propto r^{-3}\) at large radii. Because it lacks any strong central enhancement, the NFW profile leads to the lower microlensing event rate compared to the density spike, which amplifies \(\rho_{\rm DM}\) within the inner parsec and hence the PBH number density. Although the dark matter spike profile has a much higher density than the NFW profile in the innermost region of the Galaxy, the corresponding microlensing event rate only begins to increase at larger PBH masses. In the spike scenario, most PBHs orbit very close to the supermassive black hole, where the velocity dispersion reaches several hundred km\,s$^{-1}$. The microlensing event duration is approximately given by
\[
t_E \simeq \frac{R_E}{v} \propto \frac{\sqrt{M_{\rm PBH}}}{v(r)} ,
\]
meaning that for small PBH masses, the Einstein crossing time is shorter due to both low lens mass and larger velocity. Events with such short duration do not come under the ballpark of the observational detectability, and therefore do not contribute to the observable event rate. As a result, even though the number density of lenses is very high in the spike, the detectable event rate remains low at small $M_{\rm PBH}$.

At very small masses (\(M_{\rm PBH}\lesssim 10^{-2}M_\odot\)) PBHs are abundant but produce only weak lensing magnification, yielding very few detectable events. As the PBH mass increases, the lensing cross-section grows, and each PBH becomes more efficient at causing microlensing, thereby increasing the event rate. However, for sufficiently high masses (\(M_{\rm PBH}\gtrsim 10^{3}M_\odot\)) the number density of PBHs drops sharply as \(n_{\rm PBH}\propto 1/M_{\rm PBH}\), causing the total event rate to flatten or decrease. This competition between decreasing number density and increasing lensing strength produces a characteristic peak in the event rate; the location and height of the peak depend on the assumed density profile.

We assume different telescope resolutions $\sigma/R=10^{-3},10^{-4},10^{-5}$ and show that the event rate improves by order $O(10)$ for finer resolution, and a larger number of events cross the benchmark crossing time $t_e>1$.

\section{Expected exclusion on PBH parameter space}
\label{sec:exclusion}
\begin{figure}
\centering
\includegraphics[width=0.45\textwidth]{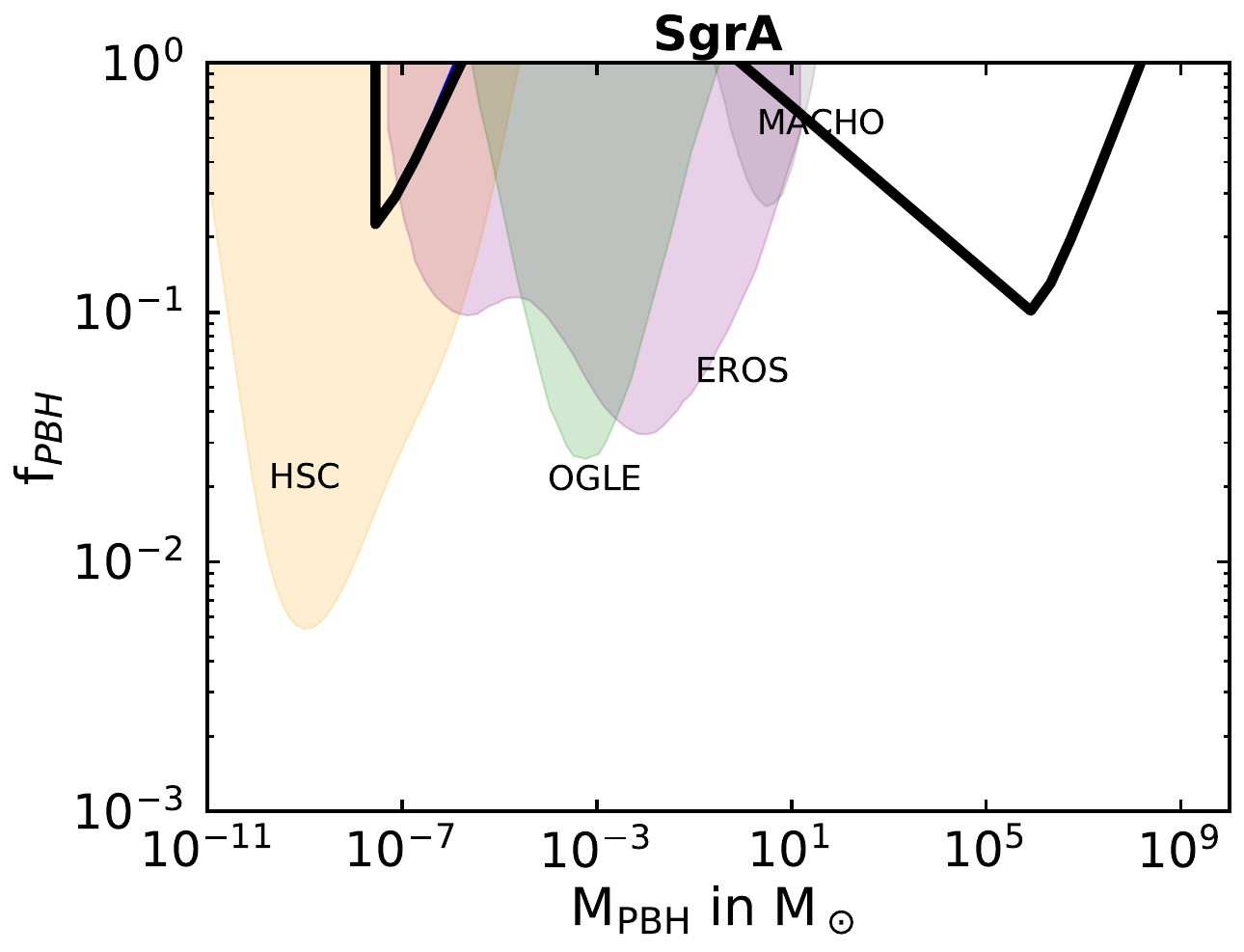}
\includegraphics[width=0.45\textwidth]{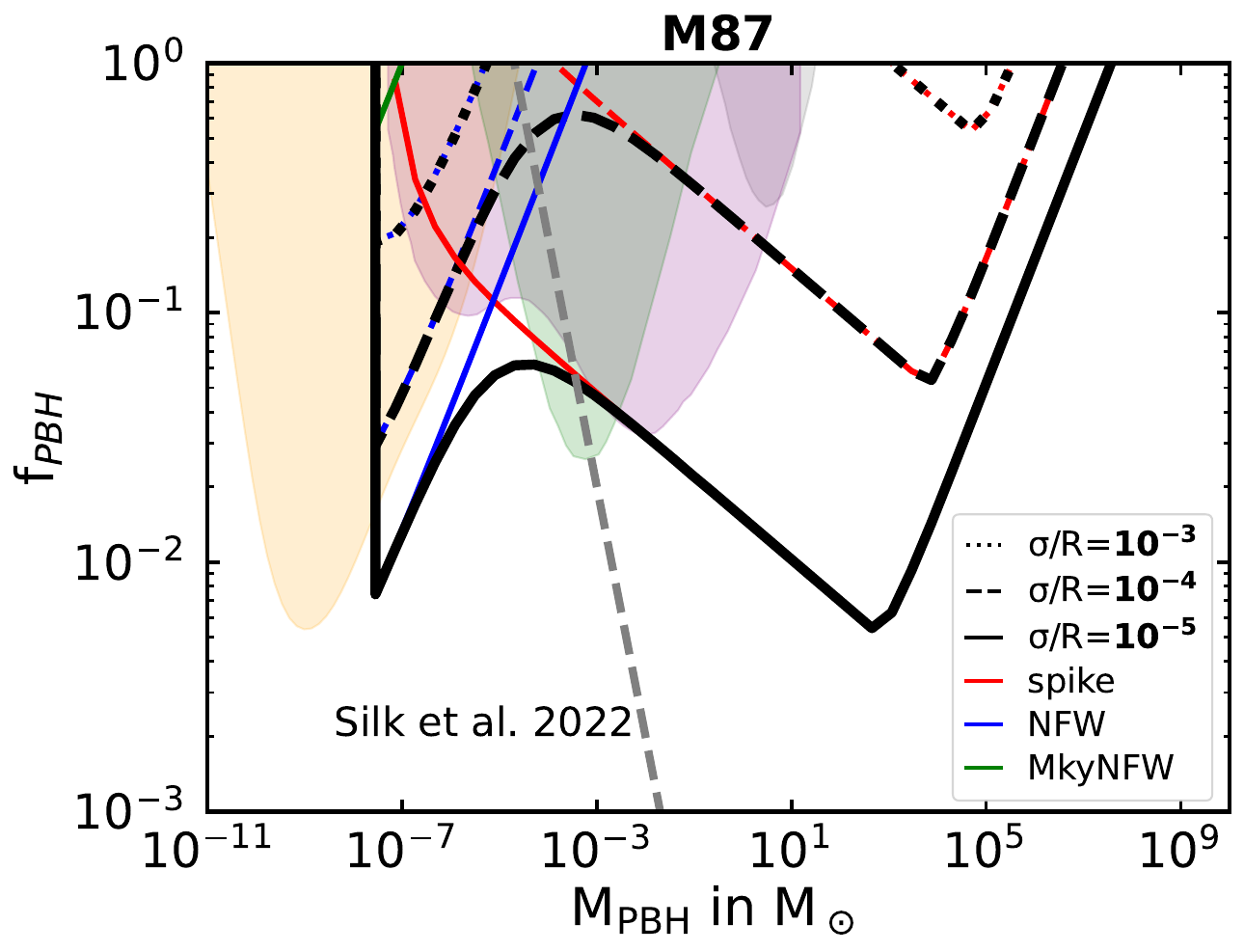}
\caption{
Constraints on the primordial black hole dark matter fraction $f_{\text{PBH}}$ as a function of PBH mass $M_{\text{PBH}}$ from shadow observations of Sgr A* (\textit{left}) and M87* (\textit{right}) with different possible resolutions $\sigma/R$. Colored lines represent constraints from different dark matter profiles: spike (red), NFW (blue), and mixed MkyNFW (green, M87 only). Line styles indicate angular resolution scenarios: dotted ($\sigma/R = 10^{-3}$), dashed ($\sigma/R = 10^{-4}$), and solid ($\sigma/R = 10^{-5}$). Thick black lines show the combined constraints from all dark matter profiles for each resolution scenario. Existing constraints from other microlensing surveys are shown in different color shades: MACHO (grey), HSC (orange), EROS (pink), and OGLE (green). For M87, the constraint from \cite{Silk:2022cck} for standard spike parameter values: $\gamma_{\rm sp}= 7/3$ (grey dashed).  Our analysis assumes 2 years of observations with a daily cadence. }
\label{fig:pbh_constraints}
\end{figure}

The microlensing constraints shown in fig.~\ref{fig:pbh_constraints} are derived under the condition of non-observation of any change in the size and shape of the photon ring between two imaging epochs separated by one year. We translate the event rate duration in fig.~\ref{fig:DifferentialEvents} into the maximum allowed PBH fractions as a function of mass. We focus on events with a characteristic duration of order one day, which typically sets the exposure time relevant for individual observation. For a monitoring cadence of one year, the probability that an event occurring during the exposure window is actually detected is suppressed by the ratio $t_E / t_{\rm cad}$, or more accurately by the Poisson probability 
\begin{equation}
    P_{\rm det} = 1 - e^{-t_E / t_{\rm cad}},
\end{equation}
where $t_E$ is the Einstein crossing time and $t_{\rm cad}$ is the cadence between successive observations. This factor suppresses the effective event rate for short-duration lenses, leading to non-detection of the event between the observing epochs. For long-duration events ($t_E \gg t_{\rm cad}$), the detection probability saturates to unity, implying that the persistent lensing can induce measurable distortion in the photon ring. Because of this reason, the bounds on the PBH fraction get stronger as we move towards increasing mass, as $t_{\rm E} \propto \sqrt{M_{\rm {PBH}}}$ is larger for large lens mass. This observation is valid for both supermassive black holes.

The colored curves in Fig.~\ref{fig:pbh_constraints}, 
maintaining the same color scheme as the event rate plots, show the individual contributions of each dark matter profile to the overall constraint: the red curve represents the bound resulting from the dark matter spike, the blue curve shows the bound from the NFW component, and finally, the green curve, present only in the M87 case, represents the constraints coming from MilkyWay NFW dark matter. The reason for showing the individual profile constraints is to illustrate which component provides the dominant constraining power at each mass. The line styles encode instrumental resolutions: dotted for currently accessible $\sigma/R = 10^{-3}$, dashed for near-future improvements to $\sigma/R = 10^{-4}$, and solid for next-generation capability at $\sigma/R = 10^{-5}$.

For both SgrA and M87, it is the dark matter spike profile provides stronger bounds on the PBH fraction. The key reason is the enhancement of the dark-matter density in the spike scenario, which increases the PBH number density and therefore the microlensing event rate $\Gamma \propto n_{\rm{PBH}}$. We start from the minimum PBH mass below which geometric optics becomes invalid ($M < M_{\text{crit}} \approx 2 \times 10^{-11} M_\odot$). 


We include several existing observational bounds from classical microlensing surveys. These limits—MACHO, EROS, OGLE-2019, and HSC—represent the most stringent constraints on PBHs across a wide range of masses \cite{MACHO2000, EROS2007,OGLE2019, HSC2019}. While the bounds are weaker, even for the spike case in SgrA, our result opens up in the higher lens mass range $M \sim 10^4-10^7~ M_\odot$. For M87, the overall event rate is larger, as seen from fig.~\ref{fig:DifferentialEvents}, hence the bounds are improved by order $O(10^{-1})$ compared to SgrA, for both NFW and spike dark matter profiles.



\section{Summary and Discussion}
\label{sec:summary}
In this work, we have investigated gravitational microlensing of black hole photon rings as a probe of PBH dark matter. Building on the formalism developed in the earlier work \cite{Verma:2023hes}, we showed that PBHs can induce time-dependent distortions in the photon ring structure of the SMBH. We analyzed the microlensing of photon rings of  Sgr~A$^*$ and M87$^*$ by PBHs distributed in dark matter configurations, including the standard NFW profile, PBH density spikes around the central SMBH, and the Milky Way foreground PBHs. While microlensing by stellar-mass objects produces minimal event rates, PBH can generate enhanced optical depths due to their larger number densities. We begin by characterizing three primary shadow observables that encode microlensing signatures: the center shift $|OC|/R$ quantifying displacement of the shadow centroid, the asymmetry $A/R$ measuring elongation perpendicular to the lens trajectory, and the fractional radius change $(\langle R_L\rangle - R)/R$. As can be seen from \cref{fig:shadobs}, it is the asymmetry that dominates
the fractional radius change by a factor of ten to a hundred across the relevant impact parameter range $\xi/\theta_E \sim 2$--$20$. This motivates our adoption of asymmetry as the detection observable throughout the subsequent event rate calculations.

For each combination of source and dark matter profile, we first evaluate the event rates as a function of PBH mass $M_{\rm PBH}$ and three representative asymmetry detection thresholds $\sigma_{A}/R = 10^{-3}$, $10^{-4}$, and $10^{-5}$. We then derive upper limits on  $f_{\rm PBH}$, by imposing the non-detection criterion of fewer than one event over the two-year baseline with daily cadence. For  Sgr~A$^*$, our most stringent constraints arise from the density spike profile. For an optimistic threshold $\sigma_A/R = 10^{-5}$, the spike model constrains $f_{\rm PBH} \lesssim 10^{-1}$ across the mass range $M_{\rm PBH} \sim 10^{1}$--$10^7 M_\odot$, with peak sensitivity near $M_{\rm PBH} \sim 10^5 -10^6 M_\odot$, where the crossing times $t_E$ optimally match the observation cadence. The constraints weaken towards higher masses where crossing times ($t_E \propto \sqrt{M}$) extend beyond the campaign duration. The NFW profile yields  weaker constraints typically $f_{\rm PBH} \lesssim 0.7$--$0.8$ across $M_{\rm PBH} \sim 10^{-8} -10^{-6} M_\odot$ with same detection threhsold $\sigma_A/R$. In contrast,  M87$^*$, the spike profile yields the most stringent bounds, reaching $f_{\rm PBH} \lesssim 10^{-2}$--$5 \times 10^{-2}$ in the intermediate-mass range $M_{\rm PBH} \sim 10$--$10^3 M_\odot$ for the optimistic threshold $\sigma_A/R = 10^{-5}$. These constraints are complementary to the existing bounds from microlensing surveys like HSC, OGLE, and extend sensitivity to PBH masses outside the reach of traditional microlensing techniques. The intermediate threshold $\sigma_A/R = 10^{-4}$ again degrades all M87* constraints by approximately an order of magnitude, with the combined model achieving $f_{\rm PBH} \lesssim 0.1$--$0.3$ in the optimal mass range, while the conservative threshold $\sigma_A/R = 10^{-3}$ approaches unity across much of the parameter space. 

Future improvements in angular resolution, baseline coverage, and temporal monitoring of the photon ring will further enhance the sensitivity of this method. The detection or non-detection of time-dependent photon-ring distortions can hence provide new constraints on PBH dark matter and demonstrate the utility of black hole imaging as a probe of dark matter nature.

\section*{Acknowledgement}
HV is supported by the funding for the Roman Galactic Exoplanet Survey Project Infrastructure Team, which is provided by the Nancy Grace Roman Space Telescope Project through the National Aeronautics and Space Administration grant 80NSSC24M0022, by The Ohio State University through the Thomas Jefferson Chair for Space Exploration endowment, and by the Vanderbilt Initiative in Data-intensive Astrophysics (VIDA). The work of PS and KC was supported by the National Science and Technology Council (NSTC) of Taiwan under grant no. MOST-113-2112-M- 007-041-MY3.

\bibliographystyle{apsrev4-1}
\bibliography{EHTmicrolensing}

\end{document}